# Stock Price Prediction using Sentiment Analysis and Deep Learning for Indian Markets

Narayana Darapaneni[1], Anwesh Reddy Paduri[2], Himank Sharma[3], Milind Manjrekar[4], Nutan Hindlekar[5], Pranali Bhagat[6], Usha Aiyer[7], Yogesh Agarwal[8]

[1] Northwestern University/Great Learning, Evanston, US
[2-8] Great Learning, Bangalore, India

anwesh@greatlearning.in

**Abstract.** Stock market prediction has been an active area of research for a considerable period. Arrival of computing, followed by Machine Learning has upgraded the speed of research as well as opened new avenues. As part of this research study, we aimed to predict the future stock movement of shares using the historical prices aided with availability of sentiment data. Two models were used as part of the exercise, LSTM was the first model with historical prices as the independent variable. Sentiment Analysis captured using Intensity Analyzer was used as the major parameter for Random Forest Model used for the second part, some macro parameters like Gold, Oil prices, USD exchange rate and Indian Govt. Securities yields were also added to the model for improved accuracy of the model. As the end product, prices of 4 stocks viz. Reliance, HDFC Bank, TCS and SBI were predicted using the aforementioned two models. The results were evaluated using RMSE metric.

## 1 Introduction

The objective of this exercise has been to predict future stock prices using Machine Learning and other Artificial Intelligence. The exercise started with a comprehensive review of available literature in this domain. Research papers as well as online sources tackling this problem were reviewed, a brief list of the same is included as part of references.

### 1.1 Literature Review

Early research on Stock Market Prediction was based on Random walk and Efficient Market Hypothesis (EMH). Numerous studies like Gallagher, Kavussanos, Butler, show that stock market prices do not follow a random walk and can be predicted to an extent [3][7][10]. Another hypothesis which is currently under survey is, whether the early indicators extracted from online sources (blogs, twitter feeds etc.) can be used to predict changes in economic and commercial indicators. Analysis of the same has been done in other fields of research, for e.g. Gruhl et al showed the correlation between online chat activity and book sales [8]. Blog sentiment assessment has been used to predict movie sales by Mishne, Glance et al [15]. Schumaker et al investigated the relations between breaking financial news and stock price changes [18]. One of the major researches in the field of stock prediction was carried out by Bollen, Mao et al 2011, where they investigated correlation between public mood and Dow Jones Industrial Index. Public moods (Happy, Calm, and Anxiety) were derived using twitter feed [2]. Chen and Lazer derived investment strategies by observing and classifying the twitter feeds [4]. Bing et al, studied the twitter feed and concluded predictability of stock based on the industry [1]. Zhang et al found high negative correlation between negative moods on social network and DJIA index [20]. Pagolu et al in their work showed a strong correlation between rise/fall of a company stock prices and the public emotions expressed on twitter. Instead of using a standard word embedding model, their work focused on developing a sentiment analyzer to categorize tweets in three categories: Positive, Negative and Neutral [16]. Mittal et al in their study tried to build a portfolio management tool using the twitter sentiment analysis. They analyzed and tested their model on DJIA. The model based on greedy strategy received feedback from sentiment analysis of the social media to predict the Buy/Sell decisions for the DJIA positions, one day in advance. Chen et al used a model established on LSTM algorithm to predict direction of stocks in Chinese Stock Exchange, in their study, they compared LSTM with Random estimation model confirming higher accuracy for the LSTM model [4]. Study by Tekin et al analysed the data of the 25 leading companies and applied various forecasting models [19]. Their studies showed a higher relevance of Random Forest technique. LSTM multi-layer perceptron (MLP) and random forest classifier are employed by Malandri et al in their Portfolio allocation model. A study of NYSE data suggests that LSTM gives better experimental results [13]. Kilimci et al, presented their study on developing an Efficient Word Embedding and Deep Learning Based Model to Forecast the Direction of Stock Exchange Market Using Twitter and Financial News Sites for Turkish Stock Exchange (BIST 100) [11]. Their study used a mix of various word embedding and Deep Learning models to arrive at the combination with highest accuracy regards prediction of the stocks. They use data labelling to classify the information as positive or negative. The data is then sent to the word embedding models like Word2Vec, FastText, GloVe to building different word embedding models to be tested with the three separate deep learning techniques viz., CNN, RNN and LSTM. The combination of Word2Vec embedding model combined with LSTM gave the highest average accuracy over the 9 stocks in consideration while using Twitter data as the base.

## 1.2    A Data sourcing, Pre-processing and EDA

The exercise started with stock related information available in the public domain. Yahoo Finance was used as the source for stock related information. The data gathered contained the regular data points referred during Stock Analysis viz., Open, Low, High, Close prices, Adjacent Close, Volume of Trade etc. Data from Jan 2007 was used as part of the EDA exercise.

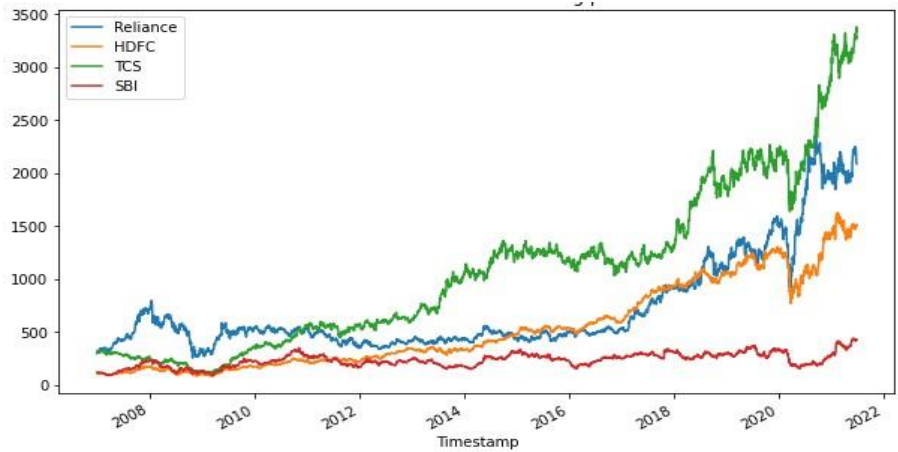

**Fig. 1.** Historical Performance of Stocks

The process of model building led us to Domain Exploration beyond the parts already covered in the Literature Survey part. Various Macro/Global Economic and other fundamental parameters were researched across domains ranging from Finance, Economy, Trade or other Core industry parameters. The focus was to finalize a list of parameters which would have a significant bearing on the stock prices. The final list of Macro parameters chosen as part of the exercise were Gold prices (it is expected that there is negative relation between gold prices and market returns), Brent Price (proxy for Fuel, Fuel prices have a significant impact of almost all economic indicators), Govt. Securities yields (Increase in bond yields puts significant pressures on the economy thus impacting Market returns) and USD-INR exchange Rate(Fluctuations in exchange rates have a significant impact on various macro parameters, it is expected to help us in better explanation and prediction of stock prices movements).

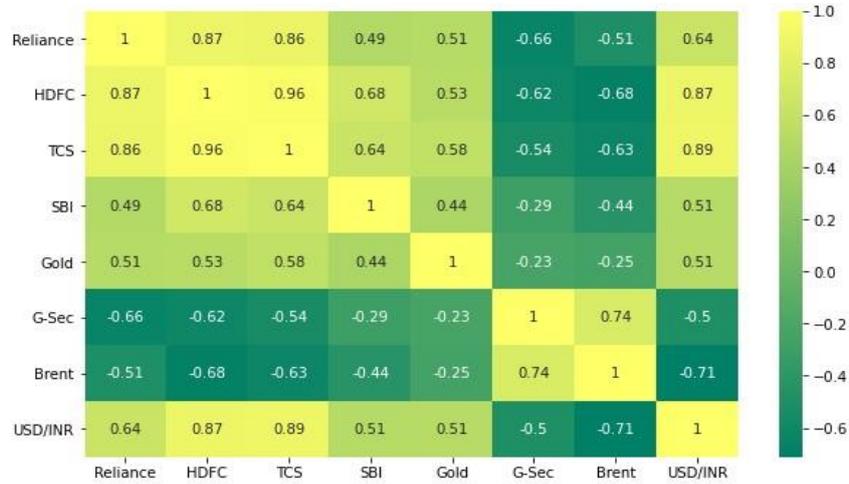

**Fig. 2.** Correlation - Macro Parameters and Stock Prices

Another significant aspect of the exercise was the usage of Sentiment Analysis. It was expected to use Tweets data as the feed for our Sentiment Analyzer but the sourcing of feeds data from twitter turned up to be an unsurmountable challenge due to rule changes in the means of accessing twitter data. As an alternative, manual approach was used to source news headlines from various publicly available data news sources like BSE, India Today, Reuters News, News18, Hindustan Times, Mint, Global Filings etc. Data of 2 years was compiled as part of the exercise ranging from 1 June 2019 to 28 June 2021. The data was available on a daily basis in the aforementioned websites, the same was compiled into an excel file with separate rows corresponding to a single news headline. Since this data was available from news websites, data pre-processing was carried out. Stop word removal exercise, special characters' removal and other standard pre-processing activities were carried out to convert the data into a format acceptable for the Sentiment Intensity Analyzer. The intensity analyzer gives 4 types of Sentiments viz., Positive, Negative, Neutral and Compound as part of the cloud. The data for prices has been considered and is being predicted on a daily level while there were several news items corresponding to a single date. Thus, all news items corresponding to a day were concatenated as one text input for the Sentiment Analyzer returning the applicable Sentiment cloud for the date. This sentiment cloud is used along with the historical Close Price and other Macro parameters mentioned above to make predictions using the Random Forest Model.

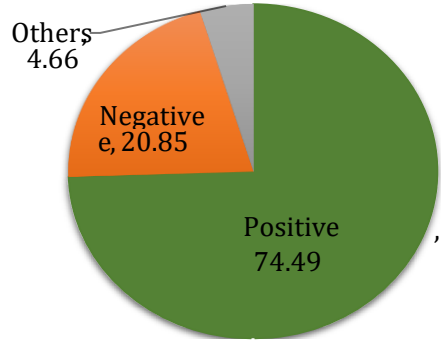

**Fig. 3.** Summary of Sentiments for Reliance

A summary of various data sources, range of data and the data sources are indicated in the table below.

**Table 1.** Data – Features, Range and Sources

| Independent Features | Indicator Used | Date Range | Data Source |
|---|---|---|---|
| **Gold Prices** | Comex 100oz Gold Price – Chicago Mercantile Exchange (Gold) | 29 Dec 2006 to 28 June 2021 | Public Domain |
| **Fuel Prices** | Brent Crude Oil Benchmark (Brent) | 29 Dec 2006 to 28 June 2021 | Public Domain |
| **Bond Yields** | 10 year Govt. of India Bond Yield (G-Sec) | 29 Dec 2006 to 28 June 2021 | Public Domain |
| **Exchange Rate** | USD-INR Exchange Rate | 29 Dec 2006 to 28 June 2021 | Public Domain |
| **Sentiment Data** | News from websites related to stocks in consideration viz., Reliance, SBI, HDFC Bank and TCS | 1 June 2019 to 28 June 2021 | Public Domain websites like BSE, India Today, Reuters News, News18, Hindustan Times, Mint, Global Filings etc. |
| **Dependent Variable Close Price** | Close prices are considered – National Stock Exchange price | 29 Dec 2006 to 28 June 2021 | Yahoo Finance |

## 2 First Section Step by step walk through of the solution

The exercise of Data pre-processing and EDA was followed with an attempt to use several Machine Learning algorithms to arrive at required error levels. Some of the algorithms like Linear Regression, KNN, Random forest Regressor, Prophet, Arima were tried. Brief description of the algorithms along with the parameters considered is indicated below:

a) Linear Regression is a linear approach to modelling the relationship between a scalar response and one or more explanatory variables (also known as dependent and independent variables). A stock's price and time-period determine the system parameters for linear regression, making the method universally applicable.

b) K-nearest neighbor's algorithm is part of a family of algorithms that allows pattern recognition. It is a type of instance-based learning, or lazy learning. Due to dependence on distance for classification, normalizing is integral to improving the accuracy. Euclidean distance metric was used in our model for prediction. Range of neighbor values from 2 to 9 were tested, Greed Search was used to arrive at the optimal n value. Cross validation for hyper-parameters was also carried out 5 times.

c) Auto regressive models are similar to a regression model but the Regressor in this case is the same dependent variable with a specific lag. Autoregressive Integrated Moving Average (ARIMA) Model converts non-stationary data to stationary data before working on it. 50 functions to be evaluated with 'lbfgs' method, AR of order 0, differencing of order 1 and MA of order 1. Seasonal Component of the model for the AR is 2, differences is 1, MA is 0, and periodicity is 12.

d) Prophet is a procedure for forecasting time series data, based on an additive model where non-linear trends are fit with yearly, weekly, and daily seasonality, plus holiday effects. It works best with time series that have strong seasonal effects and several seasons of historical data.

e) We have used linear curve to predict the daily stock price. The data frame used in Prophet required two columns-(i) "ds": to store date time series and (ii) "y": to store the corresponding values of the time series (stock values)

f) Random Forest Regressor is a meta estimator that fits a number of classifying decision trees on various sub-samples of the dataset and uses averaging to improve the predictive accuracy and control over-fitting.

g) LSTM including Bidirectional-LSTM were tried to predict stock prices through usage of Historical Close price values. Hyper parameter tuning was also carried out to arrive at the optimum results.

As indicated in the table below, LSTM is performing better as compared to other models tried as part of the study. Thus, LSTM was selected for prediction of stock prices for other companies as well viz., HDFC Bank, SBI and TCS.

**Table 2.** RMSE Values for different models

| Models | RMSE |
| --- | --- |
| LSTM | 38.19 |
| Bidirectional LSTM | 184.29 |
| Linear Regression | 1030.83 |
| Arima | 532.64 |
| KNN | 1273.05 |
| Prophet | 311.46 |
| Random Forest Regressor | 580.49 |

Sentiment analysis using News Headlines was carried out as the next step of the exercise. An attempt was made the sentiments of news data. Polarity score for each daily news i.e. Positive, Negative, Neutral & Compound values were calculated using Intensity Analyzer. The results were not meeting our expectations as higher RMSE values were observed. Further tuning of the model was carried out by adding other parameters viz., Gold, Brent, G-sec and USD/INR exchange rate. There was a marked improvement in model predictions with the addition of these parameters. RMSE values with the model were comparable to the LSTM model used above. Thus, the final solution in this regard used Random Forest Regressor with additional macro parameters for sentiment analysis.

## 3    Model Evaluation

### 3.1    LSTM Model

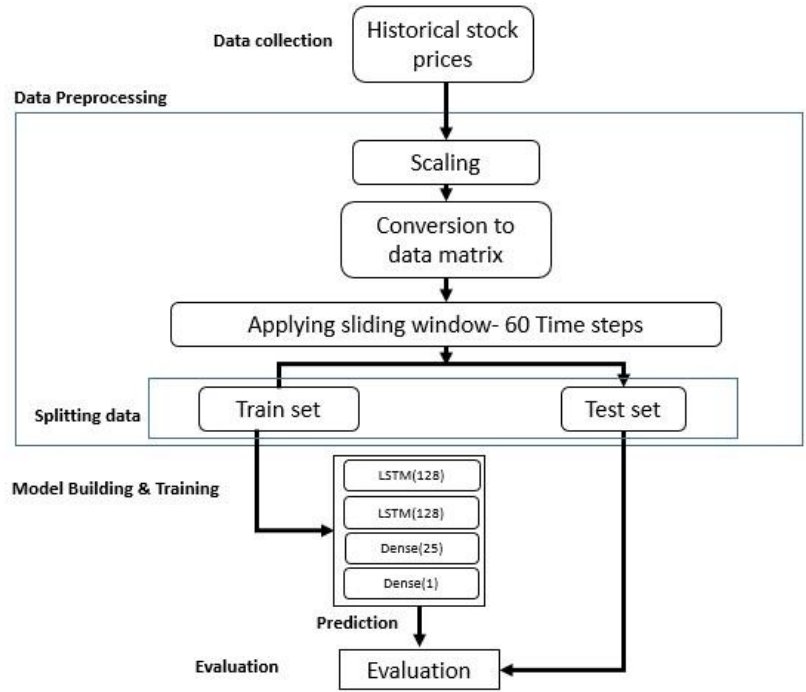

**Fig. 4.** Block diagram of stock prediction using LSTM

The purpose of this study has been to devise trading strategies based on stock price predictions, so Regression Analysis has been used to arrive at future stock price. LSTM has been the most successful in price prediction among the models we have tried. LSTM or Long-Short-Term Memory Recurrent Neural Network belongs to the family of deep learning algorithms which works on the feedback connections in its architecture. It has an advantage over traditional neural networks due to its capability to process entire sequence of data. Its architecture comprises the cell, input gate, output gate and forget gate. Data pre-processing is an important step in LSTM. Scaling of data is a process which is advisable with most models, thus LSTM also requires processing in the form of scaling. Since LSTM works on sequences using them as the base for prediction of single value. Thus, a matrix needs to be created from the date wise train data set available. The train data fed into the LSTM consists of a multi-dimensional array consisting of various instances of Dependent variable and the corresponding linked independent variable, which in our case is an array consisting historical close prices, this period is referred to as sliding window. Various ranges of 5 days to 250 days were tried for slid-ing window to ascertain best fit for the model in consideration. As part of model building, various variations of the model were tried including addition of various Dense, Dropout layers. Hyper parameter tuning was also carried out by comparing errors across different runs. Batch normalization was also tried but didn't yield any significant improvement in results. Beside the parameter tuning, Bi-Directional variation of LSTM was also attempted to get better results.

As a result of the entire model building exercise, a sliding historical window of 60 days gave the best results among the range covered. Two layers LSTM respectively with 128 and 64 neurons followed by two dense layers of 25 and 1 neurons was the final model that gave best performance among various model variations.

Since this is a regression model, standard features like accuracy % couldn't be used. Thus, RMSE was used as the quantifying parameter for evaluating the success of models being tested.

### 3.2    Random Forest - Sentiment Analysis

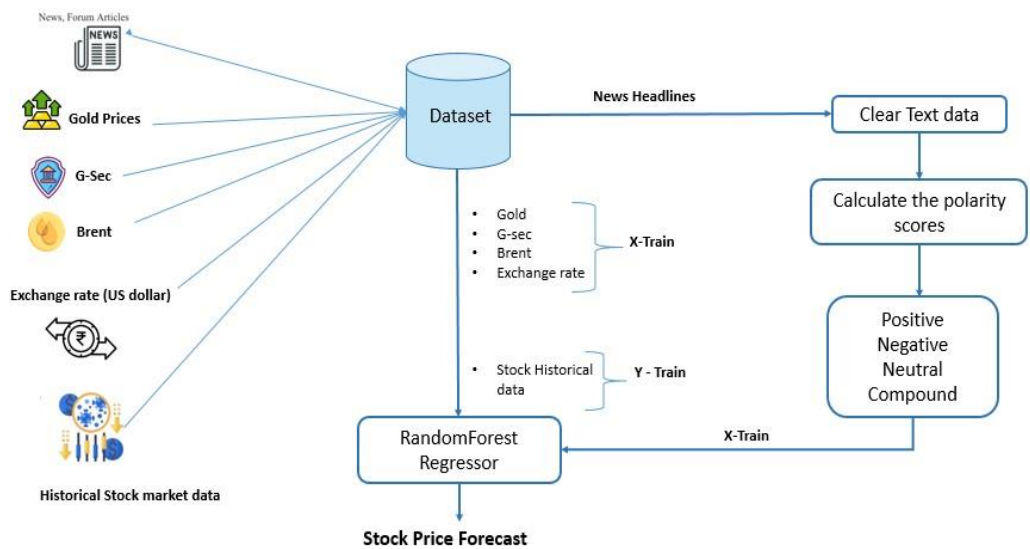

**Fig. 5.** Block diagram of sentiment analysis

The aim of this study has been to use Sentiment Analysis for prediction of Stock prices. One of the challenges with LSTM is the usage of single parameter for model building. Since LSTM could not be used for sentiment analysis.

There were two major parts of the exercise, first being daily sentiment collection and analysis and second being the building of model for the prediction of values using. As mentioned before, the first part consisted of manually sourcing data from various public domain websites. Preprocessing of data was also carried out using standard libraries in order to improve the data quality. Since there were more than one news items for a single day, all news items for a single day were concatenated to arrive at the combined news data for the day. Sentiment Intensity Analyzer, standard library was used to generate Sentiment polarity which gives 4 values corresponding to the input text. It measures the level of Positive, Negative, Neutral and Compound sentiment associated with the input text. These 4 parameters are considered as the independent features for the Sentiment part. Standard regression models were tried and after several attempts it was realized that that Random Forest Regressor was most suited for performing the analysis.

Preliminary runs of the model with using Close price and the 4 sentiment features described above produced very poor results. There was variation of 20-30% in the predicted values. Based on the feedback, domain exploration was carried out to including any other additional features. Various permutations were carried out with external features. 4 features namely GSec yield, Brent price, Gold prices, USD exchange rates were added to the model These macro parameters turned out to be quite significant in improving prediction accuracy of the model. Since this is a regression model, standard features like accuracy % couldn't be used. Thus, RMSE was used as the quantifying parameter for evaluating the success of models being tested. Both the models described above, viz., LSTM and Random Forest were used to predict the future stock prices of 4 stocks for 28th June as part of the study. Predicted values using the two models along with the actual stock prices for the day are indicated in the table below:

**Table 3.** Predicted and actual prices of Stocks

| Stocks | Actual value_28Jun | LSTM Prediction | Sentiment analysis |
|--------|--------------------|-----------------|--------------------|
| RIL | 2086 | 2106 | 2093 |
| HDFC Bank | 1508.35 | 1492 | 1504 |
| TCS | 3336.75 | 3312 | 3135 |
| SBI | 426.75 | 427 | 357 |

## 4 Visualizations

LSTM was applied on closing prices of the four stocks viz. Reliance, HDFC, TCS & SBI. Model data including the Train, Validation and Predict have been depicted in below graphs. Blue line indicates train data; Orange line indicates the validation & green line is the predicted close value for the stock. Out of total 3478 data points, 3305 data points are considered in training & rest 5% for validation which covers a span of 15 years. RMSE values for Reliance, HDFC Bank, TCS & SBI stock were 38, 33, 59 & 7 respectively. The LSTM model error is significantly lower than the error values of earlier models like Linear Regression, ARIMA & k-nearest neighbor etc.

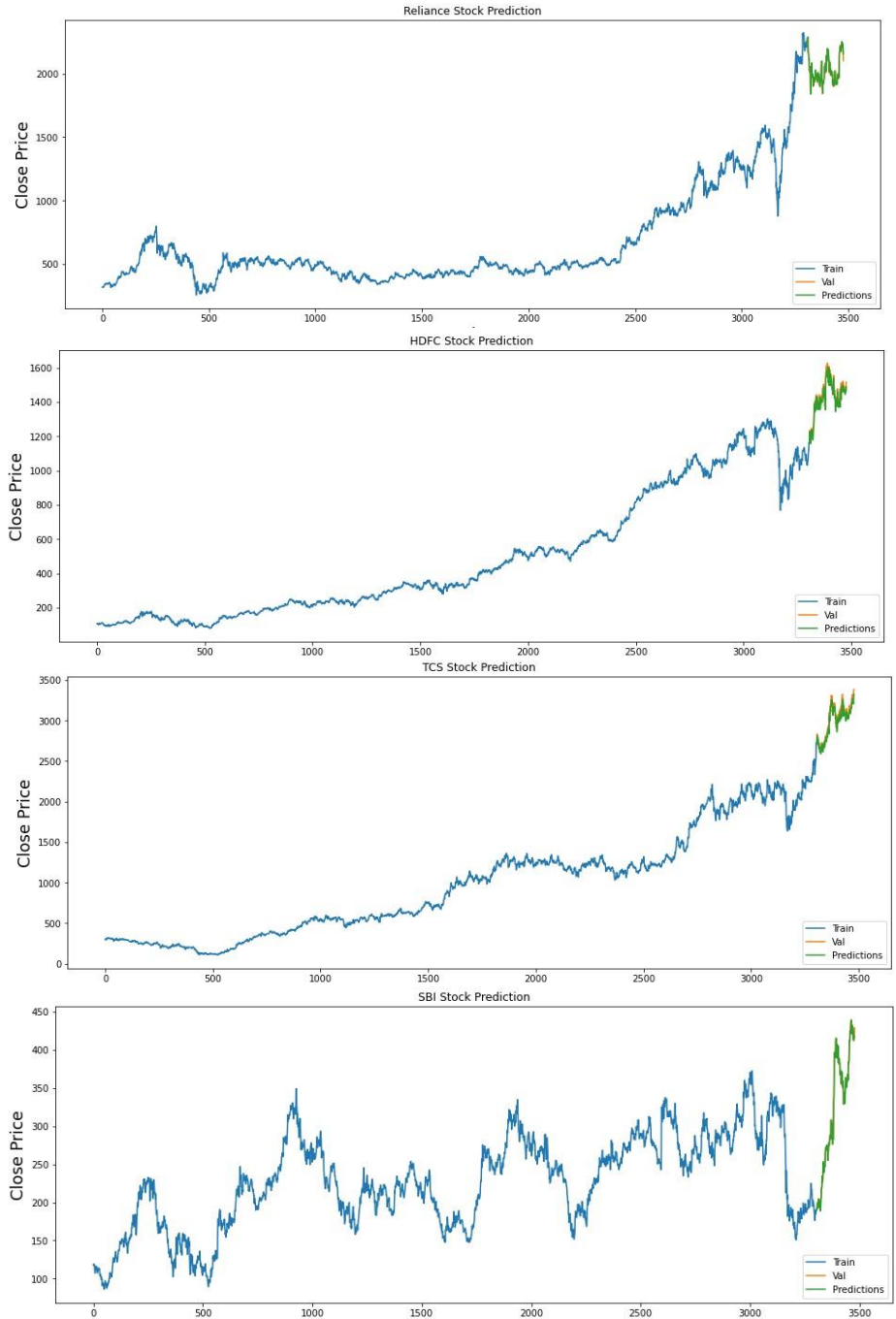

**Fig. 6.** Stock Predictions as per LSTM

Further, an attempt was made to analyze impact of daily news sentiments & external factors such as Gold, G-Sec, Brent, and INR-USD rate on stock movement using Random forest regression. Resultant output for the model has been indicated graphically below. LSTM outperforms Random forest regression, but with additional features, the Random Forest model does provide better predictions. TCS is a exception where RMSE value (139) is significantly higher than LSTM. One of the reasons could be, insufficient valid news for sentiment analysis.

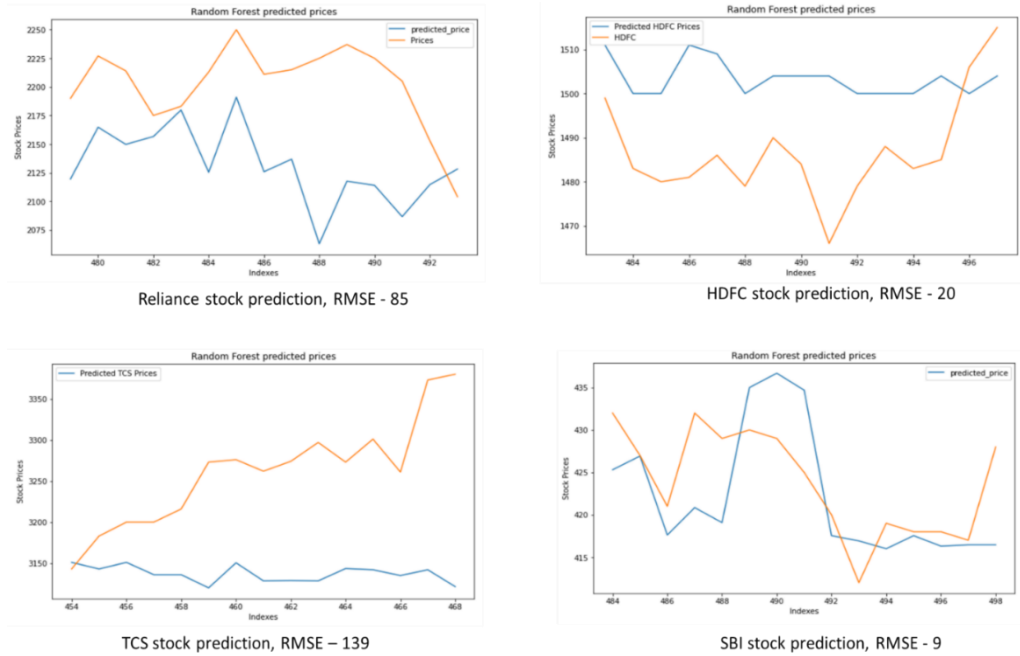

Reliance stock prediction, RMSE - 85

HDFC stock prediction, RMSE - 20

TCS stock prediction, RMSE – 139

SBI stock prediction, RMSE - 9

**Fig. 7.** Stock Predictions as per Random Forest using Sentiment Analysis

## 5 Results/Implications

This broad purpose of this exercise has been to arrive at trading strategies which could help in the real-world application of the models developed. The study could not approach to those levels due to several constraints and limitations as described above.

Two Regressor that were used viz LSTM and Random Forest have been used to predict the next day value and the RMSEs are considered as the evaluation metrics. Since this was a regression exercise prediction with certain confidence levels couldn't be achieved. Thus, a simple intuitive analysis of RMSE values was carried out to gauge the appropriateness of the values predicted by our models. The results as achieved from the models are indicated below.

**Table 4.** RMSE values and Error Ratio for Stock prices

| Stocks | LSTM RMSE | LSTM MAPE% | Senti-ment Analysis RMSE | Sentiment Analysis MAPE% |
|--------|-----------|------------|--------------------------|--------------------------|
| RIL | 38.19 | 1.36 | 85.11 | 3.41 |
| HDFC Bank | 33.14 | 1.81 | 20.49 | 1.25 |
| TCS | 59.59 | 1.60 | 139.16 | 3.76 |
| SBI | 7.89 | 1.75 | 9.65 | 1.87 |

The results above indicate that the Mean Absolute Percentage Error (MAPE) varies in the range of 1.36% to 1.81% in case of LSTM while the same varies in the range of 1.25% to 3.76% in case of Sentiment Analysis using Random Forest Model. HDFC Bank is the only stock where the sentiment analysis has worked better than LSTM. SBI is the best performing stock in both models with Sentiment Analysis and LSTM returning similar level of results. Thus, we can say that a 95% confidence level could be considered an approximate fit to explain the working of this model. The exercise hasn't been able to come up with any Trading Strategy but an attempt was made to use the models to forecast a future trend of prices instead of just predicting a single day price. The results of trend forecasting weren't satisfactory and significant changes might be needed to gain any future results in this regard.

## 6 Limitations/Closing Reflections

This exercise is a research into new approach by the authors, thus there have been some limitations due to time constraints, technical challenges and others. Some of the limitations were identified as part of the exercise, same have been briefly discussed in the subsequent paragraphs. One of the challenges on data side was regarding the news data. News data was sourced from websites using related filters. This at times lead to news items which were not as relevant to the company in question. This could lead to distortion by the way of a strong sentiment of a news item concerning an unrelated entity finding its way through the filter. One of the way of resolving this limitation is using manual annotation. Word vector techniques can also improve the data quality in such cases. Another limitation, part data and part solution is regarding computation of daily sentiment as indicated in the model description part, the approach used has been concatenation of all news data related to a single day and then calculating the sentiment cloud. This approach might have a limitation that, a strong positive/negative sentiment of a single news might be tempered down due to several neutral news thus resulting a neutral sentiment for the day. Approaches involving sentiment calculation of individual news items and then using some combination method to arrive at the resultant sentiment may yield better results.

As regards the model implementation, one of the limitations that was faced during the exercise was the use of Random Forest Regressor for Sentiment Analysis. As was observed in the first part of study, LSTM performs better than Random Forest Regressor when used for predictions using close price. Few attempts were made to try and use Multivariate LSTM to capture sentiment data along with close prices but much success couldn't be achieved regarding implementation of the same.